\documentstyle[12pt,aaspp4,flushrt]{article}
\begin{document}

\title{The Sloan Digital Sky Survey: Technical Summary}

{
\author{Donald G. York\altaffilmark{\ref{Chicago}},
J. Adelman\altaffilmark{\ref{Fermilab}},
John E. Anderson, Jr.\altaffilmark{\ref{Fermilab}},
Scott F. Anderson\altaffilmark{\ref{Washington}}, 
James Annis\altaffilmark{\ref{Fermilab}},
%John N. Bahcall\altaffilmark{\ref{IAS}},
Neta A. Bahcall\altaffilmark{\ref{Princeton}},
J. A. Bakken\altaffilmark{\ref{Fermilab}},
Robert Barkhouser\altaffilmark{\ref{JHU}},
Steven Bastian\altaffilmark{\ref{Fermilab}},
Eileen Berman\altaffilmark{\ref{Fermilab}},
William N. Boroski\altaffilmark{\ref{Fermilab}},
Steve Bracker\altaffilmark{\ref{Fermilab}},
Charlie Briegel\altaffilmark{\ref{Fermilab}},
John W. Briggs\altaffilmark{\ref{Yerkes}},
J. Brinkmann\altaffilmark{\ref{APO}}, 
%Yorke Browne\altaffilmark{\ref{Washington}},
Robert Brunner\altaffilmark{\ref{Caltech}},
Scott Burles\altaffilmark{\ref{Chicago}},
Larry Carey\altaffilmark{\ref{Washington}},
Michael A. Carr\altaffilmark{\ref{Princeton}}, 
Francisco J. Castander\altaffilmark{\ref{Chicago},\ref{Pyrenees}},
Bing Chen\altaffilmark{\ref{JHU}},
Patrick L. Colestock\altaffilmark{\ref{Fermilab}},
A. J. Connolly\altaffilmark{\ref{Pittsburgh}},
J. H. Crocker\altaffilmark{\ref{JHU}},
Istv\'an Csabai\altaffilmark{\ref{JHU},\ref{Eotvos}},
Paul C. Czarapata\altaffilmark{\ref{Fermilab}},
John Eric Davis\altaffilmark{\ref{APO}},
Mamoru Doi\altaffilmark{\ref{UTokyo}},
Tom Dombeck\altaffilmark{\ref{Chicago}},
Daniel Eisenstein\altaffilmark{\ref{IAS},\ref{Chicago},\ref{Hubble}},
Nancy Ellman\altaffilmark{\ref{Yale}},
Brian R. Elms\altaffilmark{\ref{Princeton},\ref{NAOJapan}},
Michael L. Evans\altaffilmark{\ref{Washington}},
Xiaohui Fan\altaffilmark{\ref{Princeton}},
Glenn R. Federwitz\altaffilmark{\ref{Fermilab}},
%Paul D. Feldman\altaffilmark{\ref{JHU}},
Larry Fiscelli\altaffilmark{\ref{Chicago}},
Scott Friedman\altaffilmark{\ref{JHU}},
Joshua A. Frieman\altaffilmark{\ref{Fermilab},\ref{Chicago}},
Masataka Fukugita\altaffilmark{\ref{CosmicRay},\ref{IAS}},
Bruce Gillespie\altaffilmark{\ref{APO}},
James E. Gunn\altaffilmark{\ref{Princeton}}, 
Vijay K. Gurbani\altaffilmark{\ref{Fermilab}},
Ernst de Haas\altaffilmark{\ref{Princeton}},
Merle Haldeman\altaffilmark{\ref{Fermilab}},
Frederick H. Harris\altaffilmark{\ref{Flagstaff}},
J. Hayes\altaffilmark{\ref{APO}},
Timothy M. Heckman\altaffilmark{\ref{JHU}},
G. S. Hennessy\altaffilmark{\ref{USNO}},
Robert B. Hindsley\altaffilmark{\ref{NRL}},
%Craig J. Hogan\altaffilmark{\ref{Washington}},
Scott Holm\altaffilmark{\ref{Fermilab}},
Donald J. Holmgren\altaffilmark{\ref{Fermilab}},
Chi-hao Huang\altaffilmark{\ref{Fermilab}},
Charles Hull\altaffilmark{\ref{OCIW}},
Don Husby\altaffilmark{\ref{Fermilab}},
Shin-Ichi Ichikawa\altaffilmark{\ref{NAOJapan}},
Takashi Ichikawa\altaffilmark{\ref{Tohoku}},
\v{Z}eljko Ivezi\'{c}\altaffilmark{\ref{Princeton}},
Stephen Kent\altaffilmark{\ref{Fermilab}},
Rita S.J. Kim\altaffilmark{\ref{Princeton}},
E. Kinney\altaffilmark{\ref{APO}},
Mark Klaene\altaffilmark{\ref{APO}},
A. N. Kleinman\altaffilmark{\ref{APO}},
S. Kleinman\altaffilmark{\ref{APO}},
G. R. Knapp\altaffilmark{\ref{Princeton}},
John Korienek\altaffilmark{\ref{Fermilab}},
Richard G. Kron\altaffilmark{\ref{Chicago},\ref{Fermilab}},
Peter Z. Kunszt\altaffilmark{\ref{JHU}},
D.Q. Lamb\altaffilmark{\ref{Chicago}},
B. Lee\altaffilmark{\ref{Fermilab}},
R. French Leger\altaffilmark{\ref{Washington}},
Siriluk Limmongkol\altaffilmark{\ref{Washington}},
Carl Lindenmeyer\altaffilmark{\ref{Fermilab}},
Daniel C. Long\altaffilmark{\ref{APO}},
Craig Loomis\altaffilmark{\ref{APO}},
Jon Loveday\altaffilmark{\ref{Chicago}},
Rich Lucinio\altaffilmark{\ref{APO}},
Robert H. Lupton\altaffilmark{\ref{Princeton}}, 
Bryan MacKinnon\altaffilmark{\ref{Fermilab},\ref{MerrillLynch}},
Edward J. Mannery\altaffilmark{\ref{Washington}},
P. M. Mantsch\altaffilmark{\ref{Fermilab}},
Bruce Margon\altaffilmark{\ref{Washington}},
Peregrine McGehee\altaffilmark{\ref{LosAlamos}},
Timothy A. McKay\altaffilmark{\ref{Michigan}},
Avery Meiksin\altaffilmark{\ref{Edinburgh}},
Aronne Merelli\altaffilmark{\ref{CMU}},
David G. Monet\altaffilmark{\ref{Flagstaff}},
%Jeff Morgan\altaffilmark{\ref{Fermilab}},
Jeffrey A. Munn\altaffilmark{\ref{Flagstaff}},
Vijay K.  Narayanan\altaffilmark{\ref{Princeton}},
Thomas Nash\altaffilmark{\ref{Fermilab}},
Eric Neilsen\altaffilmark{\ref{JHU}},
Rich Neswold\altaffilmark{\ref{Fermilab}},
Heidi Jo Newberg\altaffilmark{\ref{Fermilab},\ref{RPI}},
R. C. Nichol\altaffilmark{\ref{CMU}},
Tom Nicinski\altaffilmark{\ref{Fermilab},\ref{Lucent}},
Mario Nonino\altaffilmark{\ref{Trieste}},
Norio Okada\altaffilmark{\ref{NAOJapan}},
Sadanori Okamura\altaffilmark{\ref{UTokyo}},
Jeremiah P. Ostriker\altaffilmark{\ref{Princeton}},
Russell Owen\altaffilmark{\ref{Washington}},
A. George Pauls\altaffilmark{\ref{Princeton}},
John Peoples\altaffilmark{\ref{Fermilab}},
R. L. Peterson\altaffilmark{\ref{Fermilab}},
Donald Petravick\altaffilmark{\ref{Fermilab}},
Jeffrey R. Pier\altaffilmark{\ref{Flagstaff}},
Adrian Pope\altaffilmark{\ref{CMU}},
Ruth Pordes\altaffilmark{\ref{Fermilab}},
Angela Prosapio\altaffilmark{\ref{Fermilab}},
Ron Rechenmacher\altaffilmark{\ref{Fermilab}},
Thomas R. Quinn\altaffilmark{\ref{Washington}},
%Stuart Rice\altaffilmark{\ref{Chicago}},
Gordon T. Richards\altaffilmark{\ref{Chicago}},
Michael W. Richmond\altaffilmark{\ref{Rochester}},
Claudio H. Rivetta\altaffilmark{\ref{Fermilab}},
Constance M. Rockosi\altaffilmark{\ref{Chicago}}, 
%Robert Rosner\altaffilmark{\ref{Chicago}},
Kurt Ruthmansdorfer\altaffilmark{\ref{Fermilab}},
Dale Sandford\altaffilmark{\ref{Yerkes}},
David J. Schlegel\altaffilmark{\ref{Princeton}},
Donald P. Schneider\altaffilmark{\ref{PennState}}, 
Maki Sekiguchi\altaffilmark{\ref{CosmicRay}},
Gary Sergey\altaffilmark{\ref{Fermilab}},
Kazuhiro Shimasaku\altaffilmark{\ref{UTokyo}},
Walter A. Siegmund\altaffilmark{\ref{Washington}},
%Allan J. Sinisgalli\altaffilmark{\ref{Princeton}},
Stephen Smee\altaffilmark{\ref{JHU}},
J. Allyn Smith\altaffilmark{\ref{Michigan}},
S. Snedden\altaffilmark{\ref{APO}},
R. Stone\altaffilmark{\ref{Flagstaff}},
Chris Stoughton\altaffilmark{\ref{Fermilab}},
Michael A. Strauss\altaffilmark{\ref{Princeton}},
Christopher Stubbs\altaffilmark{\ref{Washington}},
Mark SubbaRao\altaffilmark{\ref{Chicago}},
Alexander S. Szalay\altaffilmark{\ref{JHU}},
Istvan Szapudi\altaffilmark{\ref{CITA}},
Gyula P. Szokoly\altaffilmark{\ref{JHU}},
Anirudda R. Thakar\altaffilmark{\ref{JHU}},
Christy Tremonti\altaffilmark{\ref{JHU}},
Douglas L. Tucker\altaffilmark{\ref{Fermilab}},
%Michael S. Turner\altaffilmark{\ref{Chicago}},
Alan Uomoto\altaffilmark{\ref{JHU}},
Dan VandenBerk\altaffilmark{\ref{Fermilab}},
Michael S. Vogeley\altaffilmark{\ref{Drexel}},
Patrick Waddell\altaffilmark{\ref{Washington}},
Shu-i Wang\altaffilmark{\ref{Chicago}},
Masaru Watanabe\altaffilmark{\ref{Kanagawa}},
David H. Weinberg\altaffilmark{\ref{Ohio}},
Brian Yanny\altaffilmark{\ref{Fermilab}},  and
Naoki Yasuda\altaffilmark{\ref{NAOJapan}}
(The SDSS Collaboration)
}

%xxxx

\newcounter{address}
\setcounter{address}{1}
\altaffiltext{\theaddress}{The University of Chicago, Astronomy \& Astrophysics
Center, 5640 S. Ellis Ave., Chicago, IL 60637
\label{Chicago}}
\addtocounter{address}{1}
\altaffiltext{\theaddress}{Fermi National Accelerator Laboratory, P.O. Box 500,
Batavia, IL 60510
\label{Fermilab}}
\addtocounter{address}{1}
\altaffiltext{\theaddress}{University of Washington, Department of Astronomy,
Box 351580, Seattle, WA 98195
\label{Washington}}
\addtocounter{address}{1}
\altaffiltext{\theaddress}{Princeton University Observatory, Princeton, NJ 08544
\label{Princeton}}
\addtocounter{address}{1}
\altaffiltext{\theaddress}{
Department of Physics and Astronomy, The Johns Hopkins University,
   3701 San Martin Drive, Baltimore, MD 21218, USA
\label{JHU}
}
\addtocounter{address}{1}
\altaffiltext{\theaddress}{Yerkes Observatory, University of Chicago, 
     373 W. Geneva St. Williams Bay, WI 53191
\label{Yerkes}}
\addtocounter{address}{1}
\altaffiltext{\theaddress}{Apache Point Observatory, P.O. Box 59,
Sunspot, NM 88349-0059
\label{APO}}
\addtocounter{address}{1}
\altaffiltext{\theaddress}{Department of Astronomy, California Institute
of Technology, Pasadena, CA 91125
\label{Caltech}}
\addtocounter{address}{1}
\altaffiltext{\theaddress}{Observatoire Midi Pyrenees, 14 ave Edouard Belin,
 Toulouse, F-31400, France
\label{Pyrenees}}
\addtocounter{address}{1}
\altaffiltext{\theaddress}{Department of Physics and Astronomy,
          University of Pittsburgh,
          Pittsburgh, PA 15260
\label{Pittsburgh}}
\addtocounter{address}{1}
\altaffiltext{\theaddress}{Department of Physics of Complex Systems,
E\"otv\"os University,
   P\'azm\'any P\'eter s\'et\'any 1/A, Budapest, H-1117, Hungary
\label{Eotvos}
}
\addtocounter{address}{1}
\altaffiltext{\theaddress}{Department of Astronomy and Research Center 
  for the Early Universe, School of Science, University of Tokyo, Hongo,
  Bunkyo, Tokyo, 113-0033, Japan
\label{UTokyo}}
\addtocounter{address}{1}
\altaffiltext{\theaddress}{Institute for Advanced Study, Olden Lane,
Princeton, NJ 08540
\label{IAS}}
\addtocounter{address}{1}
\altaffiltext{\theaddress}{Hubble Fellow
\label{Hubble}}
\addtocounter{address}{1}
\altaffiltext{\theaddress}{Department of Physics, Yale University,
PO Box 208121, New
Haven, CT 06520-8121
\label{Yale}}
\addtocounter{address}{1}
\altaffiltext{\theaddress}{National Astronomical Observatory, 2-21-1, Osawa,
Mitaka, Tokyo 181-8588, Japan
\label{NAOJapan}}
\addtocounter{address}{1}
\altaffiltext{\theaddress}{Institute for Cosmic Ray Research, University of
Tokyo, Midori, Tanashi, Tokyo 188-8502, Japan
\label{CosmicRay}}
\addtocounter{address}{1}
\altaffiltext{\theaddress}{U.S. Naval Observatory, Flagstaff Station, 
P.O. Box 1149, 
Flagstaff, AZ  86002-1149
\label{Flagstaff}}
\addtocounter{address}{1}
\altaffiltext{\theaddress}{U.S. Naval Observatory, 
3450 Massachusetts Ave., NW, 
Washington, DC  20392-5420
\label{USNO}}
\addtocounter{address}{1}
\altaffiltext{\theaddress}{Remote Sensing Division, Code 7215, Naval
  Research Laboratory, 4555 Overlook Ave. SW, Washington, DC 20375
\label{NRL}}
\addtocounter{address}{1}
\altaffiltext{\theaddress}{The Observatories of the Carnegie Institution of 
Washington, 813 Santa Barbara St, Pasadena, CA 91101
\label{OCIW}}
\addtocounter{address}{1}
\altaffiltext{\theaddress}{Astronomical Institute,
Tohoku University,
Aoba, Sendai 980-8578
Japan
\label{Tohoku}}
\addtocounter{address}{1}
\altaffiltext{\theaddress}{Merrill Lynch, 
1-1-3 Otemachi, Chiyoda-ku, Tokyo 100, Japan
\label{MerrillLynch}}
\addtocounter{address}{1}
\altaffiltext{\theaddress}{Los Alamos National Laboratory, PO Box 1663, Los 
  Alamos, NM 87545
\label{LosAlamos}}
\addtocounter{address}{1}
\altaffiltext{\theaddress}{University of Michigan, Department of Physics,
	500 East University, Ann Arbor, MI 48109
\label{Michigan}}
\addtocounter{address}{1}
\altaffiltext{\theaddress}{Royal Observatory, Edinburgh, EH9 3HJ, United
  Kingdom
\label{Edinburgh}}
%\addtocounter{address}{1}
%\altaffiltext{\theaddress}{Illinois Institute of Technology, 3300 S. Federal Street
%Chicago, Il 60616
%\label{IIT}}
\addtocounter{address}{1}
\altaffiltext{\theaddress}{Dept. of Physics, Carnegie Mellon University,
     5000 Forbes Ave., Pittsburgh, PA-15232
\label{CMU}}
\addtocounter{address}{1}
\altaffiltext{\theaddress}{Physics Department, Rensselaer Polytechnic
  Institute, SC1C25, Troy, NY 12180
\label{RPI}}
\addtocounter{address}{1}
\altaffiltext{\theaddress}{Lucent Technologies, 2000 N Naperville Rd,
Naperville, IL 60566
\label{Lucent}} 
\addtocounter{address}{1}
\altaffiltext{\theaddress}{Department of Astronomy, Osservatorio Astronomico,
  via G.B. Tiepolo 11, Trieste 34131, Italy
\label{Trieste}}
\addtocounter{address}{1}
\altaffiltext{\theaddress} {Physics Department,
            Rochester Institute of Technology,
            85 Lomb Memorial Drive,
            Rochester, NY 14623-5603
\label{Rochester}}
\addtocounter{address}{1}
\altaffiltext{\theaddress}{Department of Astronomy and Astrophysics,
The Pennsylvania State University,
University Park, PA 16802
\label{PennState}}
\addtocounter{address}{1}
\altaffiltext{\theaddress}{Canadian Institute for Theoretical Astrophysics,
University of Toronto,
60 St. George Street,
Toronto, Ontario, M5S 3H8, Canada
\label{CITA}}
\addtocounter{address}{1}
\altaffiltext{\theaddress}{Department of Physics, Drexel University,
  3141 Chestnut St., Philadelphia, PA 19104
\label{Drexel}}
\addtocounter{address}{1}
\altaffiltext{\theaddress}{
Institute of Space and Astronautical Science
Sagamihara, Kanagawa 229, Japan
\label{Kanagawa}}
\addtocounter{address}{1}
\altaffiltext{\theaddress}{Ohio State University, Dept.~of Astronomy, 140
W. 18th Ave., Columbus, OH 43210
\label{Ohio}}
}

%xxx
\newpage
\begin{abstract}
The Sloan Digital Sky Survey (SDSS) will
provide the data to support detailed investigations of the 
distribution of luminous and non-luminous matter in the Universe: 
a photometrically
and astrometrically calibrated digital imaging survey
of $\pi$ steradians above about Galactic latitude $30^{\circ}$
in five broad optical bands to a depth of 
$g' \sim 23^m$, and a spectroscopic survey of the approximately $10^6$ brightest
galaxies and $10^5$ brightest quasars found in the photometric 
object catalog produced by the imaging survey.
This paper summarizes the observational parameters and data products of the 
SDSS, and serves as an introduction to extensive
technical on-line documentation.

\end{abstract}
\keywords{instrumentation - - - cosmology: observations}

\section{Introduction}

At this writing (May 2000) the Sloan Digital Sky Survey (SDSS)
is ending its commissioning phase and beginning operations.
The purpose of this paper is to
provide a concise summary of the vital statistics of the
project, a definition of some of the terms used in the
survey and, via links to documentation in
electronic form, access to detailed descriptions of the project's 
design, hardware, and software, to serve as technical
background for the project's science papers.
The electronic material is extracted from the text (the ``Project Book'')
written to support major funding proposals, and is
available at the {\it Astronomical Journal} web site via the on-line 
version of this paper.  The official SDSS web site
({\tt http://www.sdss.org}) also provides links to the on-line Project
Book, and it can be accessed directly
at {\tt http://www.astro.princeton.edu/PBOOK/welcome.htm}. In the
discussion below we reference the chapters in the Project Book by
the last part of the URL, i.e. that following {\tt PBOOK/}. 
The versions accessible at the SDSS web sites also contain extensive
discussions and summaries of the scientific goals of the survey,
which are not included here.

The text of the on-line Project Book
was last updated in August 1997.  While there have been a number of changes
in the hardware and software described therein, the material accurately 
describes the design goals and the implementation of the major observing
subsystems.
As the project becomes operational, we will provide a series of formal
technical papers (most still in preparation),
which will describe in detail the project hardware  and software
in its actual operational state. 

Section 2 describes the Survey's objectives: the imaging depth,
sky coverage, and instrumentation. 
Section 3 summarizes the 
software and data reduction components of the SDSS and its data
products.  Section 4 reviews some recent scientific
results from the project's initial commissioning data runs, which demonstrate
the ability of the project to reach its technical goals.
All Celestial coordinates are in epoch J2000.

\section{Survey Characteristics}

The Sloan Digital Sky Survey will produce both imaging and spectroscopic
surveys over a large area of the sky.  The survey uses a dedicated 2.5 m
telescope equipped with a large format mosaic CCD camera 
to image the sky
in five optical bands, and two digital spectrographs
to obtain the spectra of about
one million galaxies and 100,000 quasars selected from the imaging data.

The SDSS calibrates its photometry using observations of a network of 
standard stars established by the United States Naval Observatory 
(USNO) 1 m 
telescope, and its astrometry using observations by an array of 
astrometric CCDs in the imaging camera.

\subsection{Telescope}

The SDSS telescope is a 2.5m f/5 modified Ritchey-Chr\'etien wide-field 
altitude-azimuth telescope
(see {\tt telescop/telescop.htm}) located at the 
Apache Point Observatory (APO), Sunspot, New Mexico ({\tt site/site.htm}).
The telescope achieves a very wide ($3^{\circ}$) distortion-free
field by the use of a large secondary mirror and
two corrector lenses.  
It is equipped with the photometric/astrometric mosaic camera
({\tt camera/camera.htm}, 
Gunn et al. 1998) and images the sky by scanning along great circles
at the sidereal rate. 
The imaging camera mounts at the Cassegrain focus.  
The telescope is also equipped with two double fiber-fed spectrographs,
permanently mounted on the image rotator,  since the spectrographs are
fiber fed. This ensures that the fibers do not flex during an exposure.
The telescope is changed from imaging mode to spectroscopic mode
by removing the imaging camera and mounting at the Cassegrain
focus a fiber plug plate, individually drilled for each field,
which feeds the spectrographs.
In survey operations, it is expected that up to nine
spectroscopic plates per night will be observed, with
the necessary plates being plugged with fibers during the day.  The telescope
mounting and enclosure allow easy access for rapid
changes between fiber plug plates and between spectroscopic
and imaging modes.  This strategy allows imaging to be done in pristine
observing conditions
(photometric sky, image size $\leq 1.5''$ FWHM) and spectroscopy to 
be done during less ideal conditions.  
All observing will be done in moonless sky.

Besides the 2.5m telescope, the SDSS makes use of three subsidiary
instruments at the site.  
The {\it Photometric Telescope} (PT) is a 0.5m telescope
equipped with a CCD camera and the SDSS filter set. Its task is to
calibrate the photometry.  
Two instruments,
a {\it seeing monitor} and a 10$\mu$m {\it cloud scanner} (Hull et
al. 1995; {\tt site/site.htm}) monitor the astronomical weather.

\subsection{Imaging Camera}

The SDSS imaging camera contains two sets of CCD arrays: the imaging array
and the astrometric arrays ({\tt camera/camera.htm}, Gunn et al. 1998).

The {\it imaging array} consists of 30 2048 
$\times$ 2048 Tektronix CCDs, placed in an
array of six columns and five rows.  The telescope scanning is aligned with
the columns.  Each row observes the sky through a different filter,
in temporal sequence $r', i', u', z'$, and $g'$.
The pixel size is 24$\mu m$
($0.396''$ on the sky). 
The imaging survey is taken in drift-scan (time-delay-and-integrate,
or TDI) mode, i.e. the camera continually sweeps the sky
in great circles, and a given
point on the sky passes through the five filters in succession. 
The effective integration time per filter
is 54.1 seconds, and the time for passage over the entire
photometric array is about 5.7 minutes ({\tt strategy/strategy.htm}; 
Gunn et al. 1998). Since the camera contains six columns of CCDs,
the result is a long {\it strip} of six {\it scanlines},
containing almost simultaneously 
observed five-color data for each of the six CCD columns.  Each CCD observes
a swath of sky $13.52'$ wide.  The CCDs are separated in the {\it row}
direction (i.e. perpendicular to the scan direction) by 
91.0 mm ($25.2'$ on the sky) center-to-center.
The observations are filled in by a second strip, offset from the 
first by 93\% of the  CCD width, to produce a filled {\it stripe},
$2.54^{\circ}$ wide,
with 8\% ($1'$) lateral overlap on each side.
Because of the camera's large field of view, 
the TDI tracking must be done
along great circles.  The Northern Galactic Cap is covered by 45
great-circle arcs (shown projected on the sky in Figures 1 and 2).

\subsection{Photometry and Photometric Calibration}

The five filters in the imaging array of the camera,
[$u', ~ g', ~ r', ~ i'$ and $z'$] have effective wavelengths of
[3590 \AA{}, 4810 \AA{}, 6230 \AA{}, 7640 \AA{} and 9060 \AA{}]
(Fukugita et al. 1996; Gunn et al. 1998).
An a priori model estimate of the telescope and camera throughputs
and of the sky brightness predicted that we would reach the 5$\sigma$ detection
limit for point sources in $1''$ seeing at [22.3, 23.3, 23.1, 22.3, 20.8]
in the ($u',g',r',i',z'$) filters, respectively, at an airmass of 1.4.
We have put formal requirements on throughput at 75\% of the values
used for the above estimation, and have demonstrated that we meet
this requirement in all bands with the possible exception of $z'$.
The sensitivity limit can be tested by finding the magnitude at which
repeat observations of a given area of sky yield 50\% reproducibility
of the objects detected.  This has been tested most thoroughly with
data taken in less than optimal seeing ($1.3''-1.6''$); nevertheless, the
50\% reproducibility level lies within a few tenths of a
magnitude of the above-quoted 5$\sigma$ detection limit in all five
bands (see Ivezi\'c et al. 2000).
The SDSS science requirements demand that photometric calibration
uncertainties for point sources be 0.02 in $r'$,
0.02 in $r'-i'$
and $g'-r'$, and 0.03 in $u'-g'$ and $i'-z'$.  To meet these
stringent requirements in both signal-to-noise ratio and
photometricity, imaging data are declared to be survey quality only if
the PT determines that the night is photometric, with a
zero-point uncertainty below 1\%, and if the seeing is better than
$1.5''$.
The imaging data saturate
at about [13,  14,  14,  14, 12] magnitudes for point sources.  

The magnitude scale is on the $AB_{\nu}$ system (Oke 1969,
unpublished), which was updated to the $AB_{79}$ system by Oke \& Gunn (1983)
and to $AB_{95}$ by Fukugita et al. (1996). The magnitudes m are
related to flux density $f$ by m$\sim sinh^{-1}(f)$ 
rather than logarithmically (see Lupton, Gunn \& Szalay 1999  and Fan et al.
1999).
This definition is essentially identical to the logarithmic magnitude
at signal-to-noise ratios greater than about 5 and is well behaved for
low and even zero and negative flux densities.

The calibration and definition
of the magnitude system is carried out by the 
USNO 1 m telescope and the 0.5m PT.
The SDSS photometry is placed 
on the $AB_{\nu}$ system using
three {\it fundamental standards} ($\rm BD+17^{\circ}4708$, 
$\rm BD+26^{\circ}2606$, and $\rm BD+21^{\circ}609$), whose magnitude scale
is as defined by Fukugita et al. (1996);
a set of 157 {\it primary standards}, which are calibrated by the 
above fundamental standards using the USNO 1m telescope, and which
cover the whole range of right ascension and enable the calibration
system to be made self-consistent; and a set of {\it secondary
calibration patches} lying across the imaging stripes, containing
stars fainter than $14^m$ whose magnitudes are calibrated 
by the PT with respect
to those of the primary standards and which transfer that calibration to the
imaging survey.  The locations of these patches on the survey stripes
are shown in Figure 1.
On nights when the 2.5 m is observing, the PT
observes primary standard stars to provide the 
atmospheric extinction coefficients
over the night and to confirm that the night is photometric.
The standard star network is described 
in {\tt photcal/photcal.htm} --- note that
the telescope described there has now been replaced 
by the 0.5m PT.

\subsection{Astrometric Calibration}

The camera also contains leading and trailing {\it astrometric arrays} ---
narrow (128$\times$2048),
neutral-density-filtered, $r'$-filtered CCDs covering the
entire width of the camera.  These arrays can measure 
objects in the magnitude range
$r' \sim$ 8.5 - 16.8, i.e. they cover the dynamic range between the standard
astrometric catalog stars and the brightest unsaturated stars in the
photometric array. The astrometric calibration is thereby referenced to
the fundamental astrometric catalogues (see 
{\tt astrom/astrom.htm}), using the Hipparcos and Tycho Catalogues 
(ESA 1997) and specially observed equatorial fields (Stone et al. 1999).
Comparison with positions from the FIRST (Becker et al. 1995)
and 2MASS (Skrutskie 1999) catalogues shows that the rms astrometric accuracy 
is currently better than 150 milliarcseconds
(mas) in each coordinate.

\subsection{Imaging Survey: North Galactic Cap}

The imaging survey covers about 10,000 contiguous square degrees in the 
Northern Galactic Cap.  This area lies basically above Galactic latitude 
$30^{\circ}$, but its footprint is adjusted slightly to lie within the minimum
of the Galactic extinction contours (Schlegel, Finkbeiner \& Davis 1998),
resulting in an elliptical region.  The region is
centered at $\alpha ~ = ~ 12^h ~ 20^m$, $\delta ~ = ~
+32.5^{\circ}$.  The minor axis is at an angle $20^{\circ}$
East of North
with extent $\pm 55^{\circ}$.  The major axis is a great
circle perpendicular to the minor axis with extent
$\pm 65^{\circ}$.  The survey footprint with the location of the
stripes is shown in Figure 2 --- see 
{\tt strategy/strategy.htm} for details.

\subsection{Imaging Survey: The South Galactic Cap}

In the South Galactic Cap, three stripes will be observed, one
along the Celestial Equator and the other two north and south
of the equator (see Figure 2).  
The {\it equatorial stripe} ($\alpha$ = $20.7^h$ to 
$4^h$, $\delta$ = $0^{\circ}$) will be observed repeatedly, 
both to find variable objects
and, when co-added, to reach magnitude limits about $2^m$ deeper than 
the Northern imaging survey.

The other two stripes will cover great circles lying between 
$\alpha,~\delta$
of ($20.7^h$, -$5.8^{\circ} \rightarrow 4.0^h$, 
-$5.8^{\circ}$)
and ($22.4^h, 8.7^{\circ} \rightarrow 2.3^h$, $13.2^{\circ}$). 

\subsection{The Spectroscopic Survey}

Objects are detected in the imaging survey,
classified as point source or extended, and 
measured, by the image analysis software (see below).  
These imaging data are used to select in a uniform way
different classes of objects
whose spectra will be taken.  The final details of this {\it
target selection} will be described once the survey is well underway;
the criteria discussed here are likely to be very 
close to those finally used.

Two samples of {\it galaxies} are selected from the objects
classified as ``extended''.  About $9 \times 10^5$
galaxies will be selected to have Petrosian (1976) magnitudes
$r'_P ~ \leq$ 17.7. 
Galaxies with a mean  $r'$ band surface brightness within the half light radius
fainter than 24 magnitudes/$\rm arc~second^2$
will be removed, since spectroscopic
observations are unlikely to produce a redshift. For illustrative purposes,
a simulation of a slice of the
SDSS redshift survey is shown in Figure  3
(from Colley et al. 2000).  Galaxies in this CDM simulation are `selected'
by the SDSS selection criteria.  
As Figure 3 demonstrates,
the SDSS volume is large enough to contain a statistically significant
sample of the largest structures predicted.  

The second sample, of approximately $10^5$ galaxies, exploits the 
characteristic very red color
and high metallicity (producing strong absorption lines) of the 
most luminous galaxies: the 
``Brightest Cluster Galaxies'' or ``Bright Red Galaxies'' (BRGs);
redshifts can be well measured with the SDSS spectra for these galaxies
to about $r'$ = 19.5.
Galaxies located at the dynamical centers of nearby dense clusters often have
these properties. Reasonably accurate photometric redshifts
(Connolly et al. 1995) can be determined for these
galaxies, allowing the selection by magnitude and $g'r'i'$ color
of an essentially {\it distance
limited} sample of the highest-density regions of the Universe to a redshift of
about 0.45 (see Figure 4 for a simulation).

With their power-law continua and the influence of Lyman-$\alpha$
emission and the 
Lyman-$\alpha$ forest, {\it quasars} have $u'g'r'i'z'$ colors quite distinct
from those of the vastly more numerous stars over most of their
redshift range (Fan 1999).  Thus about $1.5 \times 10^5$ {\it quasar
candidates} are selected for spectroscopic observations as outliers
from the stellar locus (cf., Krisciunas et al.  1998; Lenz et
al. 1998; Newberg et al. 1999; Figure 5 below) in
color-color space.  At the cost of some loss of efficiency, selection
is allowed closer to the stellar locus around z = 2.8, where quasar
colors approach those of early F and late A stars (Newberg \& Yanny
1997; Fan 1999).  Some further regions of color-color space outside
the main part of the stellar locus where quasars are very rarely found
are also excluded, including the regions containing M dwarf-white dwarf pairs,
early A stars, and white dwarfs (see Figure 5).  The
SDSS will compile a sample of quasars brighter than $i' \approx$
19 at $z <$ 3.0; at redshifts between 3.0 and about 5.2, the limiting
magnitude will be about $i'$ = 20.
Objects are also required to be
point sources, except in the region of color-color space where
low-redshift quasars are expected to be found.  Stellar
objects brighter than $i'$ =20 which are FIRST sources (Becker, White
and Helfand 1995) are also selected.  Based on early spectroscopy, we
estimate that roughly 65\% of our quasar candidates are genuine
quasars; comparison with samples of known quasars indicates that our
completeness is of order 90\%.  

In all cases, the magnitudes of the objects are corrected for Galactic 
extinction before selection, using extinction in the SDSS bands calculated from
the reddening map of Schlegel, Finkbeiner \& Davis (1998).  Objects
are then selected to have a 
magnitude limit {\it outside} the Galaxy.
If this correction were not made, the systematic effects of Galactic extinction
over the survey area would overwhelm the statistical uncertainties in the
SDSS data set.
After the imaging and 
spectroscopic survey is completed in a given part of the sky, the reddening
and extinction will be recalculated using internal standards
extracted from the imaging data.  The SDSS plans to use a variety of 
extinction probes,
including very hot halo subdwarfs, halo turnoff stars, and elliptical
galaxies whose intrinsic colors can be estimated from their line indices.

Together with various classes of calibration stars and fibers which
observe blank sky to measure the sky spectrum, the selected
galaxies and quasars are mapped onto the sky, and `tiled', i.e.
their location on a $3^{\circ}$ diameter plug plate determined ({\tt 
tiling/tiling.htm}).  The centers of the tiles are adjusted to provide
closer coverage of regions of high galactic surface density, to make
the spectroscopic coverage optimally uniform.
Excess fibers are allocated to several classes of rare or peculiar
objects (for example objects which are positionally matched with
ROSAT sources, or those whose parameters lie outside any known range --
these are {\it serendipitous} objects) and to samples of stars.
The spectra are observed, 640 at a time (with a total integration time
of 45 - 60 minutes depending on observing conditions) 
using a pair of fiber-fed double spectrographs
({\tt spectro/spectro.htm}).  The wavelength 
coverage of the spectrographs is continuous from
about 3800 \AA{} to 9200 \AA{},
and  the wavelength resolution, 
$\lambda/\delta\lambda$, is 1800 (Uomoto et al. 1999).  
The fibers are located at
the focal plane via plug plates constructed for each area of sky.
The fiber diameter is 0.2 mm ($3''$ on the sky), 
and adjacent fibers cannot be located
more closely than $55''$ on the sky.  Both members of a pair of objects
closer than this separation can be observed spectroscopically
if they are located in the overlapping regions of adjacent tiles.

Tests of the redshift accuracy using observations of stars 
in M67 whose radial velocities are accurately known (Mathieu et al. 1986)
show that the SDSS radial velocity measurements  for stars have a 
scatter of about 3.5 $\rm km~s^{-1}$. 

\section{Software and Data Products}

The operational software is described in {\tt datasys/datasys.htm}.
The data are obtained using the Data Acquisition (DA) system
at APO (Petravick et al. 1994)  
and recorded on DLT tape. The imaging
data consist of full images from all CCDs of the imaging array, cut-outs
of detected objects from the astrometric array, and bookkeeping 
information.  These tapes are shipped to Fermilab by express courier 
and the data are automatically reduced through an interoperating set 
of software pipelines operating in a common computing environment.

The {\it photometric pipeline} 
reduces the imaging data; it corrects the data for data defects
(interpolation over bad columns and bleed trails, finding and interpolating
over `cosmic rays', etc), calculates
overscan (bias),
sky and flat field values, calculates the point spread functions
(psf) as a function of time and location on the CCD array, 
finds objects, combines the data from the
five bands, carries out simple model fits to the images of each
object, deblends overlapping
objects, and measures positions,
magnitudes (including psf and Petrosian magnitudes)
and shape parameters.  The photometric pipeline uses
position calibration information from the astrometric array
reduced through the {\it astrometric pipeline}
and photometric calibration data from the 
photometric telescope (reduced through the {\it photometric telescope
pipeline}).  Final calibrations are applied
by the {\it final calibration pipeline}, which allows refinements
in the positional and photometric calibration 
as the survey progresses.  The photometric pipeline
is extensively tested using repeat observations, examination of the outputs,
observations of regions of the sky previously observed
by other telescopes (HST fields,
for example) and a set of simulations, described
in detail in {\tt simul/simul.htm}.  For an example of the repeatability of 
SDSS photometry over several timescales, see Ivezi\'c et al. (2000).
These repeat observations show that the mean errors (for point sources)
are about $0.03^m$
to $20^m$, increasing to about $0.05^m$ at $21^m$ and to $0.12^m$
at $22^m$.  These observed errors are in good agreement with those quoted
by the photometric pipeline.  They apply only to the $g', r'$ and $i'$
bands -- in the less sensitive $u'$ and $z'$ bands, the errors at the
bright end are about the same as those in $g'r'i'$, but increase to 
$0.05^m$ at $20^m$ and $0.12^m$ at $21^m$.

The outputs, together with 
all the observing and processing information, are loaded into 
the {\it operational data base} 
which is the central collection of scientific and bookkeeping data 
used to run the survey. To select the spectroscopic targets, objects
are run through the {\it target selection pipeline} 
and flagged if
they meet the spectroscopic selection criteria for a particular type of 
object. The criteria for the primary objects (quasars, galaxies and BRGs)
will not be changed once the survey is underway.  Those for serendipitous
objects and samples of interesting stars  can be changed throughout the
survey.
A given object can in principle receive several target flags.
The selected objects are tiled as described above, plug
plates are drilled, and the spectroscopic observations are
made.  The spectroscopic data are automatically reduced by the
{\it spectroscopic pipeline}, which extracts,
corrects and calibrates
the spectra, determines the spectral types, and measures the redshifts.
The reduced spectra are then stored in the operational data base.
The contents of the operational data base are copied at regular 
intervals into the
{\it science data base} for 
retrieval and scientific analysis (see {\tt appsoft/appsoft.htm}).
The science data base is indexed in a hierarchical manner: the data
and other information are linked into `containers'
that  can be divided and subdivided as necessary,
to define easily searchable regions
with approximately the same data content. This hierarchical scheme is
consistent with those being adopted by other large surveys, to 
allow cross referencing of multiple surveys.  The science data base also
incorporates a set of query tools and is designed for easy portability.

The photometric data products of the SDSS include: a {\it catalog} of all 
detected objects, with measured positions, magnitudes, shape parameters,
model fits and processing flags;
{\it atlas images} (i.e. cutouts from the imaging data in all five
bands) of all detected objects and of objects from the FIRST and ROSAT
catalogs;
a $4 \times 4$ {binned image} of the 
corrected images with the objects removed: and a {\it mask}
of the areas of sky not processed (because of saturated stars,
for example) and of corrected pixels (e.g. those from which cosmic rays
were removed). The atlas images are sized to enclose
the area occupied by each object plus the PSF width, or the object
size given in the ROSAT or FIRST catalogues. The photometric outputs are
described  in
{\tt http://www.astro.princeton.edu/SDSS/photo.html}.
The data base will also contain the calibrated 1D
{\it spectra}, 
the derived {\it redshift} and {\it spectral type}, and the bookkeeping
information related to the spectroscopic observations.  In addition,
the positions of astrometric calibration stars measured 
by the astrometric pipeline and the magnitudes of the
faint photometric standards
measured by the photometric telescope pipeline will be published
at regular intervals.

\section{Early Science from the SDSS Commissioning Data}

The goal of the SDSS is to provide the data necessary for studies of the 
large scale structure of the Universe on a wide range of scales. 
The imaging survey should detect $\sim
5 \times 10^7$ galaxies, $\sim 10^6$ quasars and $\sim 8 \times 10^7$
stars to the survey limits. These photometric data, via photometric
redshifts and various statistical techniques such as the angular
correlation function, support studies of large scale structure well
past the limit of the spectroscopic survey.  On even larger scales, 
information on structure will come from quasars.  

The science justification for the SDSS is discussed in several conference
papers (e.g. Gunn \& Weinberg 1995; Fukugita 1998; Margon 1999). 
The Project Book science sections
can be accessed at
{\tt http://www.astro.princeton.edu/PBOOK/science/science.htm}.  Much 
of the science for which the SDSS was built, the study of large
scale structure, will come when the survey is complete, but the initial
test data have already led to significant scientific discoveries in many fields.
In this section, we show examples of
the first test data and some 
initial results. To date (May 2000),
the SDSS has obtained test imaging data for some 2000 square degrees
of sky and about 20,000 spectra.  Examples of these data are shown in
Figures 5 (sample color-color and color-magnitude diagrams of point-source
objects), 6 (sample spectra) and 7
(a composite color image of a piece of the sky which 
contains the cluster Abell 267),

Fischer et al. (2000) have
detected the signature of the weak lensing of background galaxies by
foreground galaxies, allowing the halos and total masses of the
foreground galaxies to be measured. 

The searches by Fan et al.
(1999a,b; 2000a,c), Schneider et al. (2000) and Zheng et al.
(2000) have greatly increased
the number of known high redshift (z$>$3.6) quasars and include
several quasars with z $>$ 5.
Fan et al. (1999b) have found the first example of a new kind of 
quasar: a high redshift object with a featureless spectrum and without the
radio emission
and polarization characteristics of BL Lac objects.  The redshift
for this object ($z$ = 4.6) is found from the Lyman-$\alpha$
forest absorption in the spectrum.

Some 150 distant probable RR Lyrae stars have been found in the
Galactic halo, enabling the halo stellar density to be mapped; the
distribution may have located the edge of the halo at approximately 60 kpc
(Ivezi\'{c} et al. 2000).  The distribution of RR Lyrae stars and other
horizontal branch stars is very clumped, showing the presence of
possible tidal streamers in the halo (Ivezi\'c et al. 2000;
Yanny et al. 2000).
Margon et al. (1999) describe the discovery of faint high latitude carbon
stars in the SDSS data. 

Strauss et al. (1999), Schneider et al. (2000), Fan et al. (2000b), 
Tsvetanov et al. (2000), Pier et al. (2000) and Leggett et al. (2000)
report the discovery of a 
number of very low mass stars or substellar objects, those of type
`L' or `T', including the first field methane (`T') dwarfs and
the first stars of spectral type intermediate between `L' and `T'.  The 
detection rate to date shows that the SDSS is likely to identify several 
thousand  L and T dwarfs.  These objects are found to occupy very
distinct regions of color-color and color-magnitude space, which
will enable the completeness of the samples to be well characterized.

Measurements of the psf diameter variations and the image wander allow
variations in the turbulence in the Earth's atmosphere to be tracked.
These data demonstrate the presence of anomalous refraction on scales
at least as large as the $\rm 2.3^{\circ}$ field of view of the camera
(Pier et al. 1999).

Of course the most exciting possibility for any large survey which probes new
regions of sensitivity or wavelength is the discovery of exceedingly
rare or entirely new classes of objects.  The SDSS has already found a 
number of very unusual objects; the nature of some of these remains
unknown (Fan et al. 1999c).
These and other investigations in progress show the promise of
SDSS for greatly advancing astronomical work in fields ranging
from the behavior of the Earth's atmosphere to structure on the
scale of the horizon of the Universe.

\acknowledgements
The Sloan Digital Sky Survey (SDSS) is a joint project of The
University of Chicago, Fermilab, the Institute for Advanced Study, the
Japan Participation Group, The Johns Hopkins University, the
Max-Planck-Institute for Astronomy, Princeton University, the United
States Naval Observatory, and the University of Washington.  Apache
Point Observatory, site of the SDSS, is operated by the Astrophysical
Research Consortium.  Funding for the project has been provided by the
Alfred P. Sloan Foundation, the SDSS member institutions, the National
Aeronautics and Space Administration, the National Science Foundation,
the U.S. Department of Energy, and Monbusho.  The official SDSS web site
is {\tt www.sdss.org}.

\newpage
\begin{figure}
\vspace{-0.5cm}
\epsfysize=500pt \epsfbox{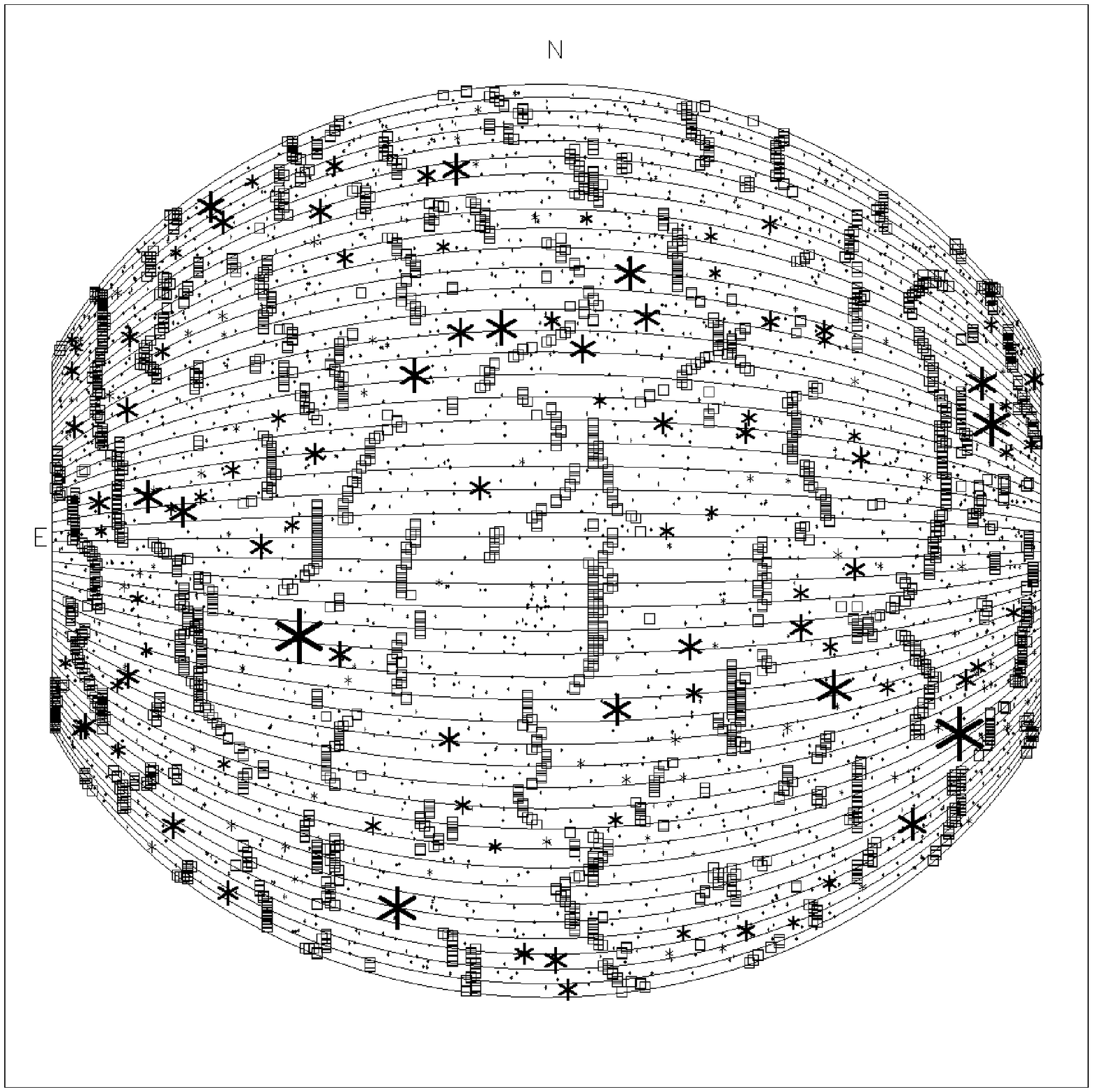}
\vspace{0.5cm}
Figure 1.
Projection on the sky of the northern SDSS survey area.
The positions of the Yale Bright Star Catalogue stars are shown.  The largest
symbols are stars of $0^m$ and the smallest stars of $5^m$ -- $7^m$.
The secondary calibration patches are shown by squares.
\end{figure}

\newpage
\begin{figure}
\vspace{-0.5cm}
\epsfysize=500pt \epsfbox{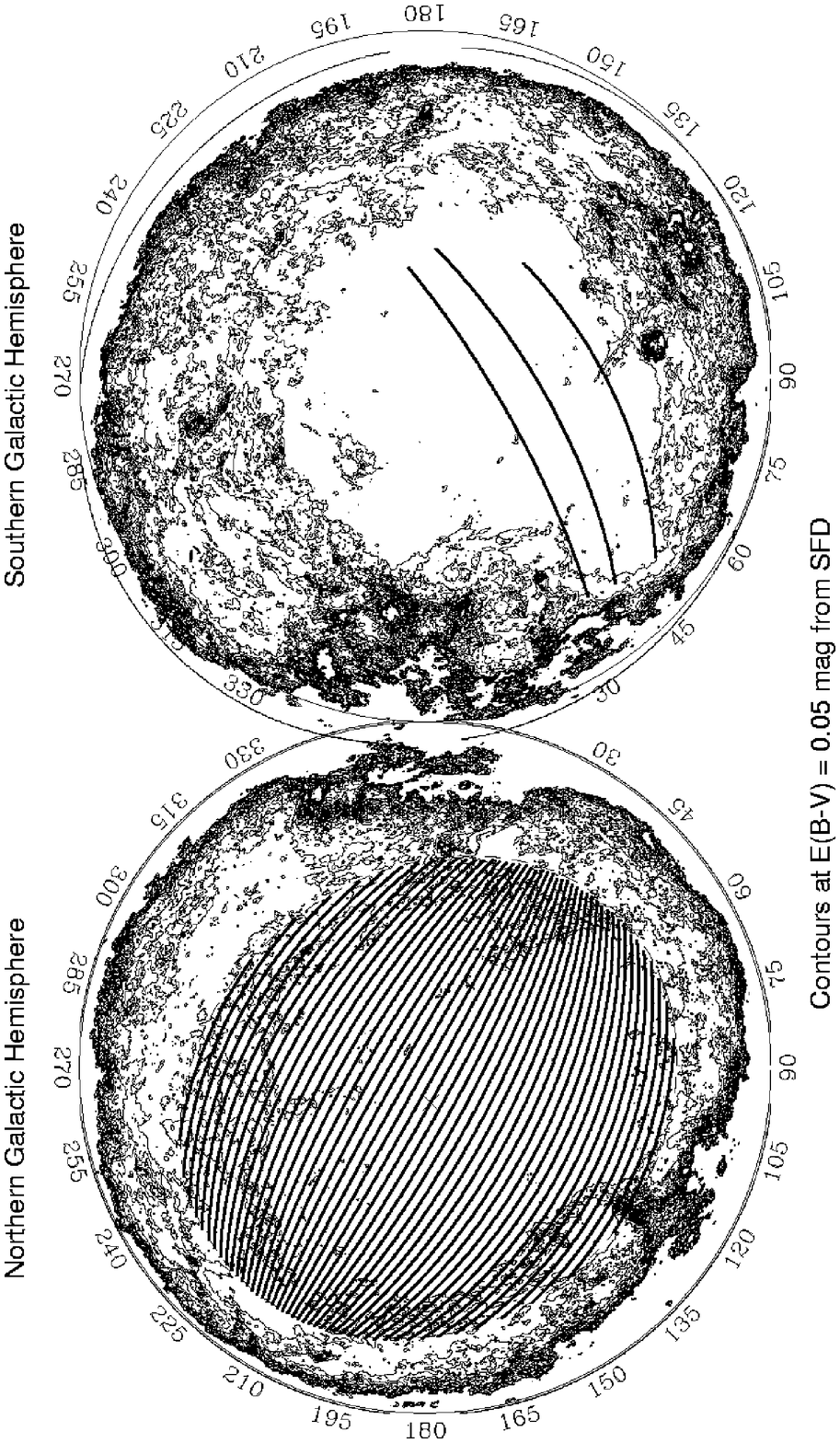}
\vspace{0.5cm}
Figure 2.
Projection on the sky (Galactic coordinates) of the
Northern and Southern SDSS surveys.  The lines show the individual
stripes to be
scanned by the imaging camera.  These are overlaid on the extinction contours
of Schlegel, Finkbeiner and Davis (1998).  The Survey pole is marked by
the `X'.
\end{figure}

\newpage
\begin{figure}
\vspace{-0.5cm}
\epsfysize=500pt \epsfbox{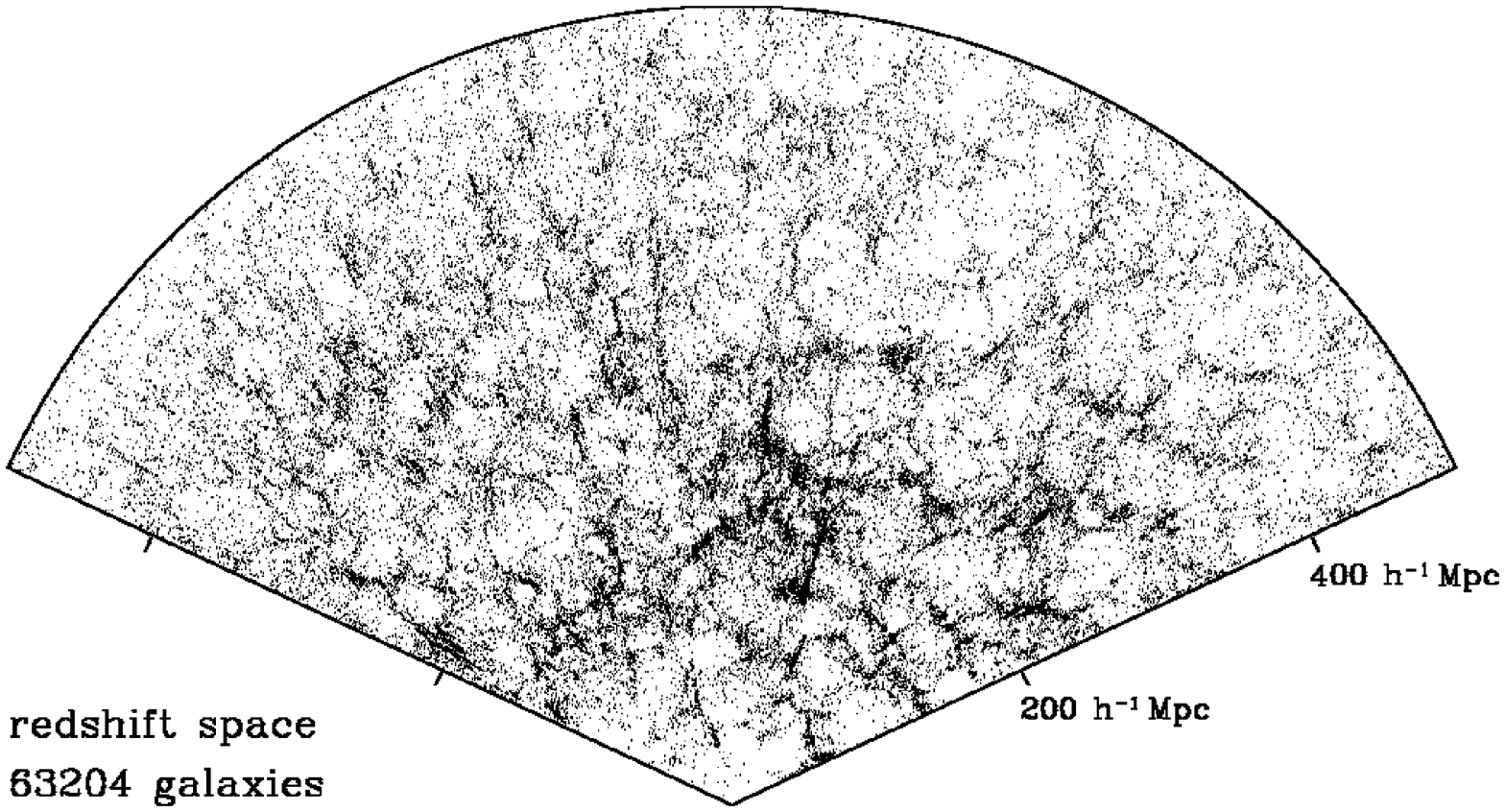}
\vspace{0.5cm}
Figure 3.
A six-degree wide slice of the Simulated Sloan Digital
Sky Survey (from Colley et al. 2000), showing about 1/20 of the survey.
\end{figure}

%\newpage
%\begin{figure}
%\vspace{-0.5cm}
%\epsfysize=500pt \epsfbox{gk4.ps}
%\vspace{0.5cm}
%Figure 4.
%Simulated redshift distribution in a $6^{\circ}$
%slice of the SDSS.  Small  dots: main galaxy sample (cf. Figure 3).
%Large dots: the BRG sample, showing about 1/30 of the survey.
%\end{figure}

\newpage
\begin{figure}
\vspace{-0.5cm}
\epsfysize=500pt \epsfbox{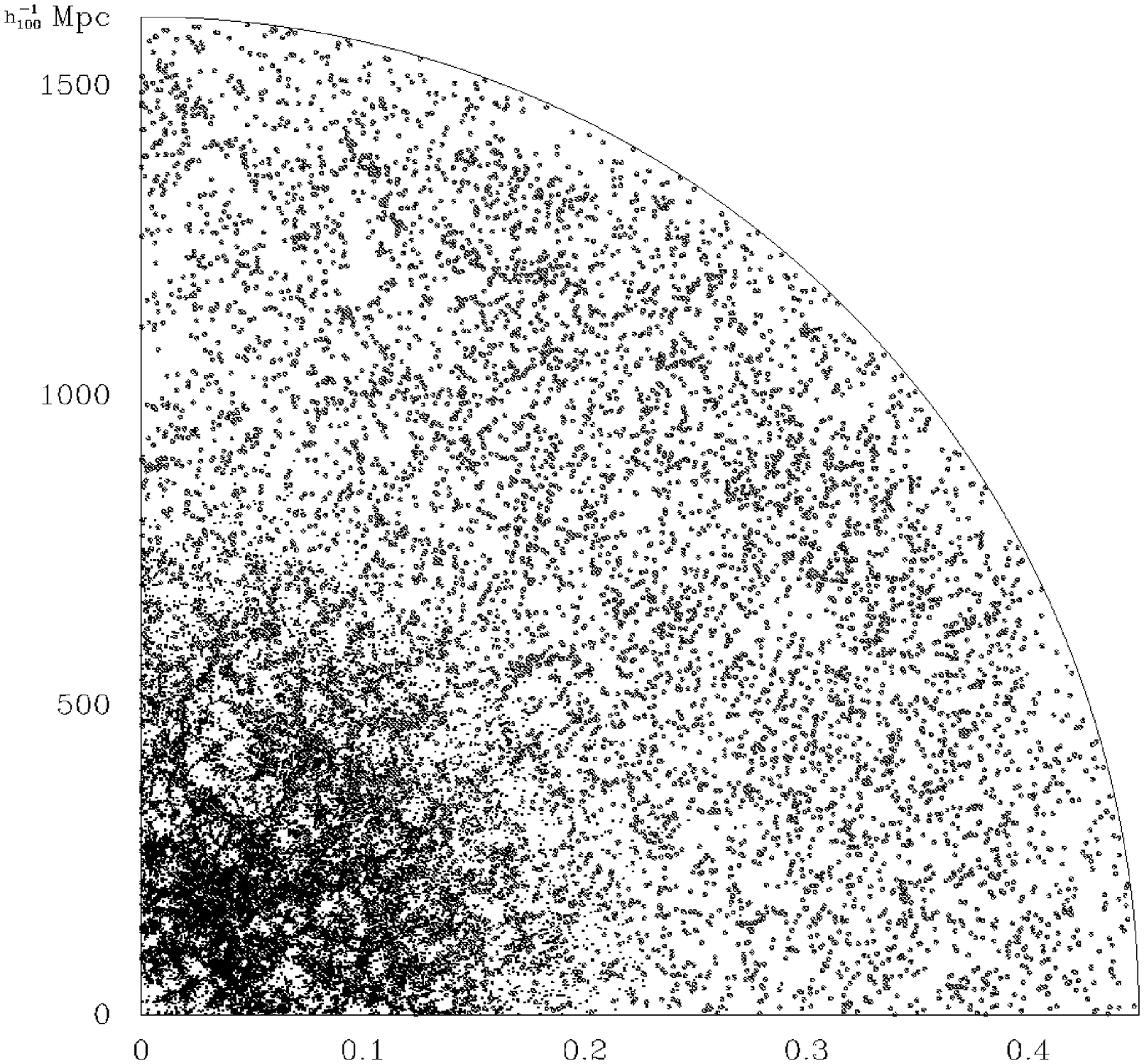}
\vspace{0.5cm}
Figure 4.
Simulated redshift distribution in a $6^{\circ}$
slice of the SDSS.  Small  dots: main galaxy sample (cf. Figure 3).
Large dots: the BRG sample, showing about 1/30 of the survey.
\end{figure}

\newpage
\begin{figure}
\vspace{-0.5cm}
\epsfysize=500pt \epsfbox{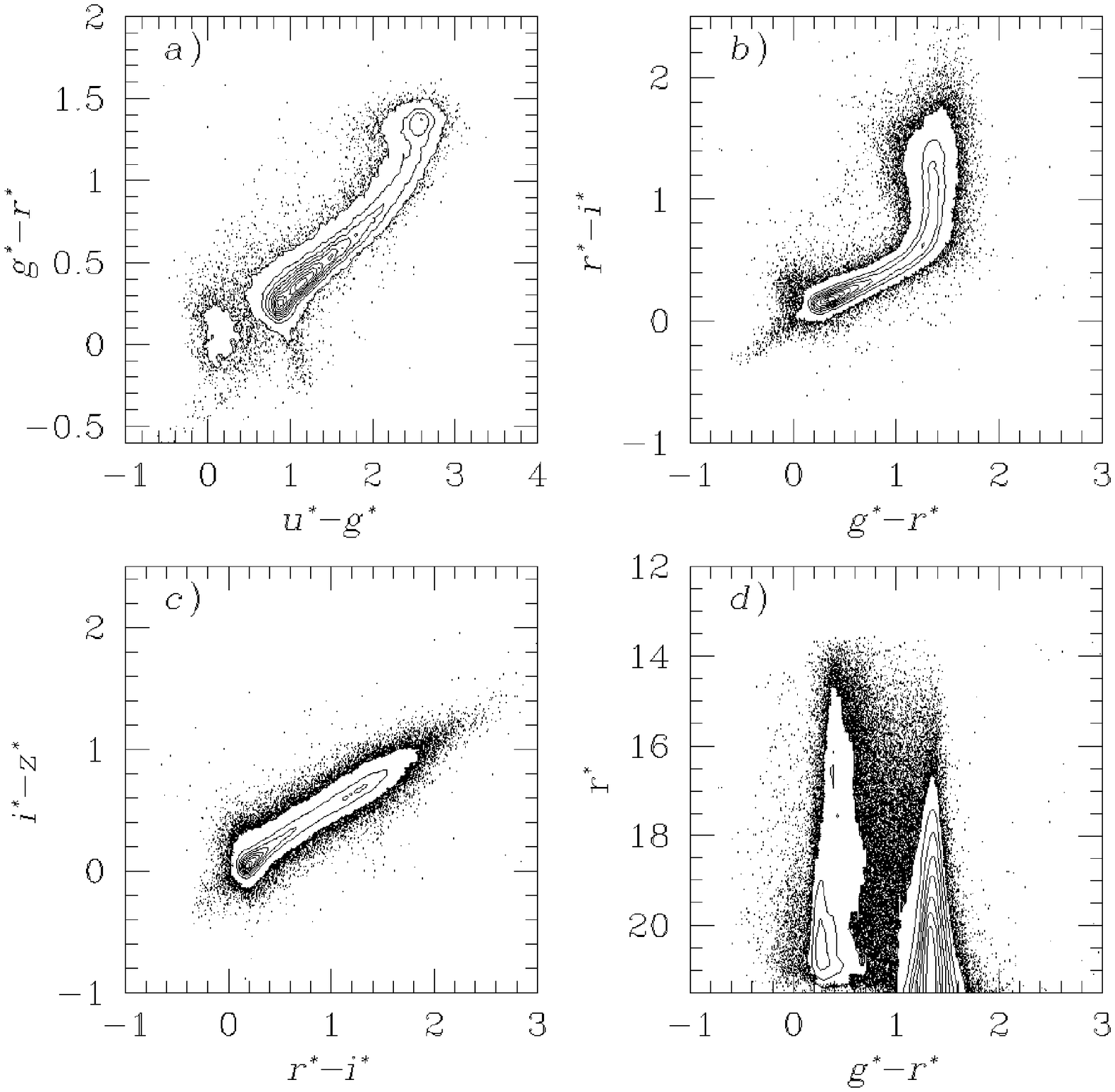}
\vspace{0.5cm}
Figure 5.
Color-color and color-magnitude plots of 
about 117,000 point sources brighter than $21^m$ at $i^*$
and detected at greater than 5$\sigma$ in each band
from 25 square degrees of
SDSS imaging data, reduced by the photometric 
pipeline (the $i^*$ designation is used for preliminary SDSS photometry). 
The contours are drawn at intervals of 10\% of the peak
density of points.  The redder stars extend to fainter magnitudes than 
do the bluer 
stars, due to the $i^*$ limit.
\end{figure}

\newpage
\begin{figure}
\vspace{-0.5cm}
\epsfysize=400pt \epsfbox{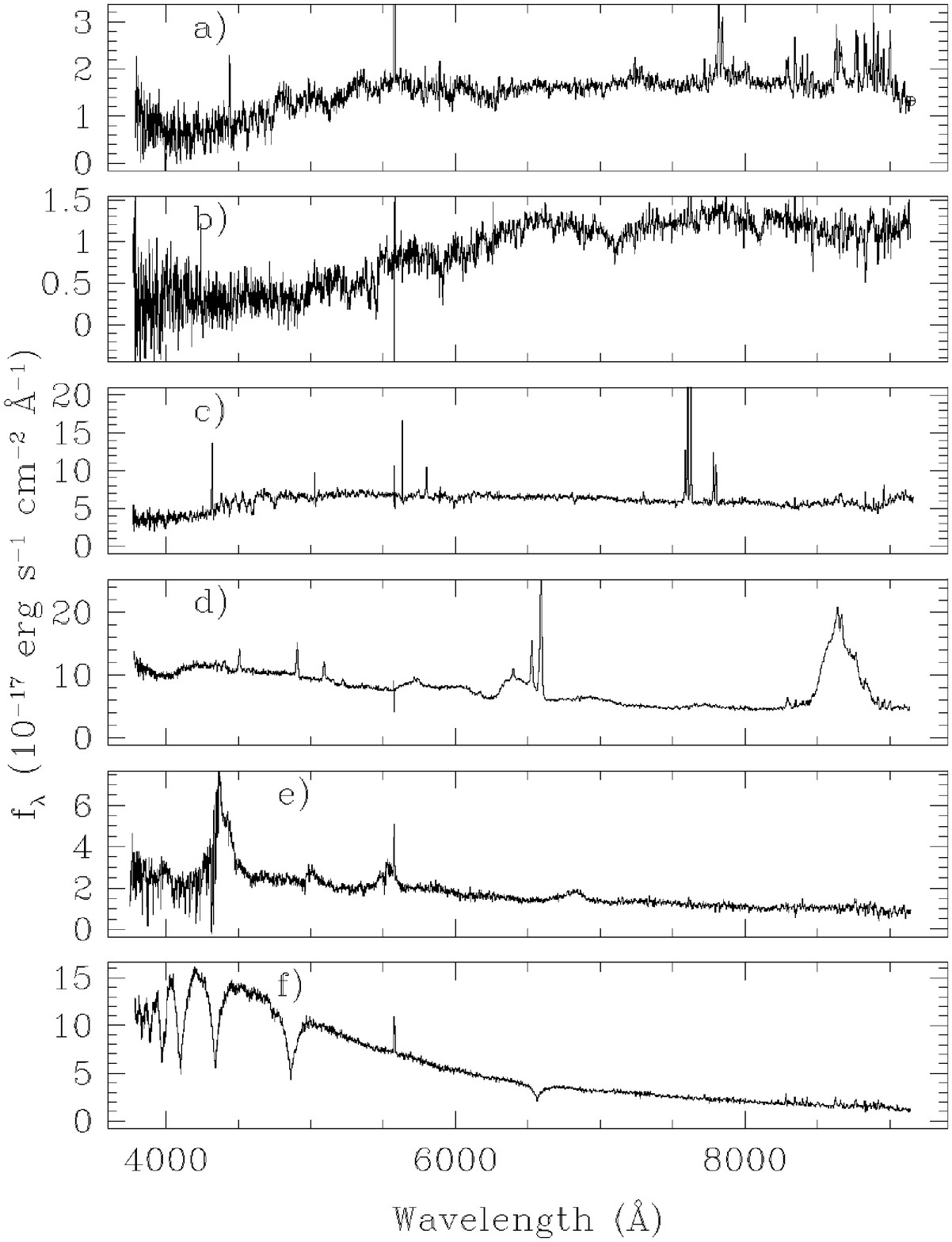}
\vspace{0.5cm}
Figure 6.
Representative SDSS spectra taken from a single spectroscopic
plate observed on 4 October 1999 for a total of one hour of integration time,
processed by the SDSS spectroscopic pipeline. For display purposes,
all spectra have been smoothed with a 3-pixel boxcar function.
All spectra show significant residuals due to the strong sky line at 5577\AA.
The objects depicted are:
a.  An $r^*_P$ = 18.00 galaxy; $z = 0.1913$.  This object is
slightly fainter than the main galaxy target selection limit.  Note
the H$\alpha$/[N II] emission at $\sim 7800$\AA, [OII] emission at
$\sim 4450$\AA, and  Ca II H and K absorption at $\sim 4700$\AA.
b. An $r^*_P$ = 19.41 galaxy, $z = 0.3735$.  This object is
close to the photometric limit of the Bright Red Galaxy sample.  The H
and K lines are particularly strong. 
c. A star-forming galaxy with $r^*_P$ = 16.88, at $z =
0.1582$.
d. A $z = 0.3162$ quasar, with $r^*_{psf}$ = 16.67. Note the unusual
profile shape of the Balmer lines. This quasar is 
LBQS 0004+0036 (Morris et al. 1991).
e. A $z = 2.575$ quasar with $r^*_{psf}$ = 19.04; 
note the resolution of the Lyman-$\alpha$
forest.  This quasar was discovered
by Berger \& Fringant (1985). 
f. A hot white dwarf, with $r_{psf}^*$ = 18.09.
\end{figure}

\newpage
\begin{figure}
\vspace{-0.5cm}
\epsfysize=300pt \epsfbox{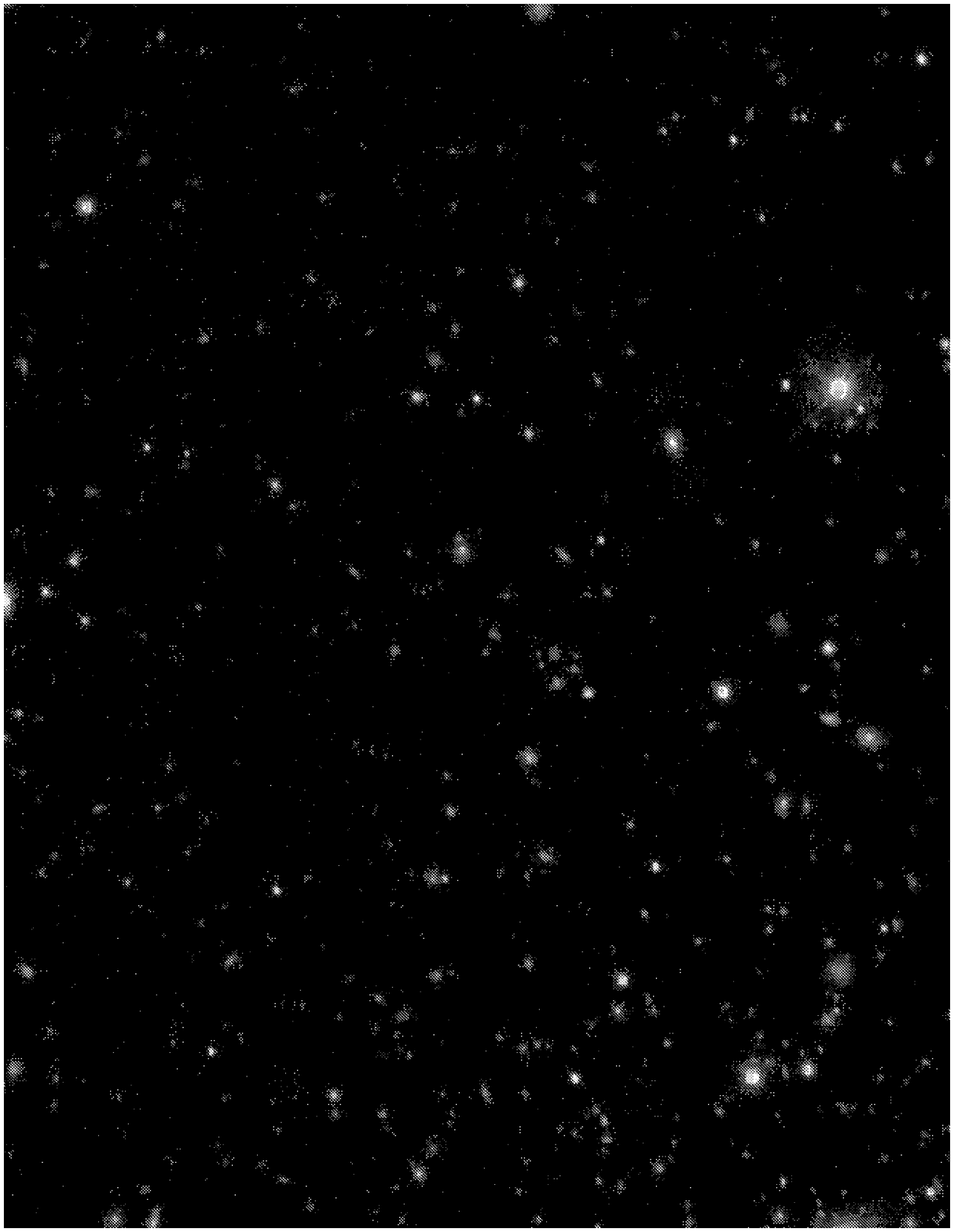}
\vspace{0.5cm}
Figure 7.
A sample frame ($13' \times 9'$) from the SDSS
imaging commissioning data.  The image, a  color
composite made from the $g', r'$,
and $i'$ data, shows a field containing the distant cluster
Abell 267 ($\alpha ~ = ~ 01^h~52^m~41.0^s, ~ \delta ~ = ~
+01^{\circ}~00'~24.7''$, redshift $z$ = 0.23, Crawford et al. 1999); 
this is the cluster of galaxies  with
yellow colors in the lower center of the frame.  The frame also contains,
in the upper center, the nearby cluster
RX J0153.2+0102, estimated redshift $\sim 0.07$ (Bade et al. 1998)
($\alpha ~ = ~ 01^h~53^m~15.15^s, ~ \delta ~ = ~ +01^{\circ}~02'~18.8''$).  
The psf (optics plus seeing) was about $\rm 1.6''$.
Right ascension increases from bottom to top of the frame, declination from
left to right.
\end{figure}

\end{document}